\documentclass[%
reprint, showpacs,
amsmath,amssymb,
aps,
prapplied,
]{revtex4-2}
\usepackage[utf8]{inputenc}
\usepackage{latexsym}
\usepackage{amssymb}
\usepackage{amsmath}
\usepackage{amsfonts}
\usepackage{cancel}
\usepackage{graphicx}
\usepackage{titlesec}
\usepackage{xcolor}

\titleformat{\paragraph}
{\normalfont\normalsize\bfseries}{\theparagraph}{1em}{}
\titlespacing*{\paragraph}
{0pt}{3.25ex plus 1ex minus .2ex}{1.5ex plus .2ex}

\begin{document}

\title{\large \textbf{Optimization-free Approach for Analog Filter Design through Spatial and Temporal Soft Switching of the Dielectric Constant}}

\author{Ohad Silbiger and Yakir Hadad}
\email{hadady@eng.tau.ac.il}
\affiliation{School of Electrical Engineering, Tel-Aviv University, Ramat-Aviv, Tel-Aviv, Israel, 69978}

\date{\today}

\begin{abstract}
The inverse-scattering problem of an inhomogeneous material has been of interest for many years, and was generally addressed with various optimization techniques. In this paper, we suggest an optimization-free method for solving the inverse-scattering problem of a one-dimensional inhomogeneous medium and use this to demonstrate the design of desired reflection frequency response. In addition, we derive the governing equation of an analog problem - a time-dependent homogeneous medium and use the same technique to design a temporal switching profile for the design of frequency response in k-space.
\end{abstract}

\maketitle

\section{Introduction}
The behavior of electromagnetic waves in spatially \cite{Collin, Walker1946} and temporally \cite{Morgenthaler1958, Felsen1970, Fante1971} varying media has been studied for a long time. The propagation of waves in a homogeneous time-dependent media shares similarities with steady-state wave propagation in inhomogeneous media, and has gained a lot of interest in recent years, due to the added degree of freedom the time variance allows. Time-dependent metamaterials have been reported to give rise to unique wave phenomena, such as nonreciprocal wave transfer \cite{Shaltout2015, Hadad2016, Sounas2017} temporal photonic crystals \cite{Zurita2009, Martinez2016}, wideband impedance matching \cite{Shlivinski2018}, electromagnetic isolators \cite{Lira2012, Taravati2017}, unitary energy transfer \cite{Mazor2021} and exotic wave reflection phenomena \cite{Bacot2016, Pacheco2021}. In many cases, however, it is of interest to solve the inverse problem. For example, if the material's properties are unknown, or if specific behavior of the waves is desired. Most previous studies addressed the inverse scattering problem in space, i.e., in the presence of non-homogeneities, and used frequency-domain reflection information to reconstruct the profile, through iterative methods \cite{Tijhuis1984, Wang1989} or mathematical approximants of the reflection coefficient \cite{Tabbara1979, Mazzarella1991}, for example. In this paper we suggest a direct, optimization-free technique, to reconstruct the dielectric profile of a one-dimensional  material  from the reflection coefficient spectrum. Furthermore, we derive an analog model of a homogeneous but temporally-varying medium that shows very similar behavior, and use the same approach to reconstruct the temporal profile from k-space measurements of the reflection. Lastly, we show how each of these models can be used to design a desired frequency response. The temporal switching adds a degree of freedom, that together with spatial non-homogeneity can allow high flexibility for the design of analog computing devices.


\section{Theory}

\subsection{Reflection in a heterogeneous medium}

When an electromagnetic wave travels in a medium with varying dielectric and magnetic properties, the non-homogeneity will cause reflections. For example, a time-harmonic TEM wave propagating in a tapered transmission line along the $z$ axis will be comprised of a superposition of forward and backward propagating waves (see Fig \ref{fig:1_TL}(a)). The total voltage on the line can be represented using the reflection coefficient $\Gamma(z)$:
\begin{equation}
    V(z) = V^+(z) e^{j(\omega t - \beta(z) z)} \left[1 + \Gamma(z) \right]
\end{equation}
The reflection coefficient is subject to Riccati's equation \cite{Collin}:
\begin{equation} \label{eq:riccati_space}
    \frac{d\Gamma}{dz} = 2j\beta \Gamma - \frac{1}{2} (1 - \Gamma^2) \frac{d (\ln Z)}{dz}
\end{equation}
Where $\beta$ is the propagation constant in the transmission line: $\beta(z) = \omega \sqrt{\mu(z)\epsilon(z)}$ and $Z$ is it's characteristic impedance: $Z(z) = \sqrt{\mu(z)/\epsilon(z)}$.

\subsection{Reflection in a temporally varying medium}

Riccati's equation governs the reflection coefficient of a time-harmonic TEM wave in an inhomogeneous medium. In this section we derive the governing equation for an analog problem - a space-harmonic TEM wave in a homogeneous medium. Consider a TEM wave propagating in an infinite transmission line containing a homogeneous medium with permittivity $\epsilon_1$ and permeability $\mu_1$:
\begin{equation}
    V = V_1 \, e^{j(kz - \omega t)} , \qquad
    I = \frac{V_1}{Z_1} \, e^{j(kz - \omega t)}
\end{equation}
\begin{figure}[h]
    \includegraphics{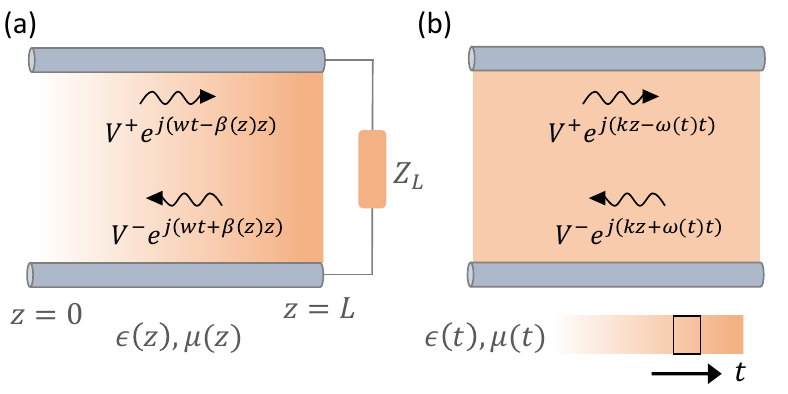}
    \centering
    \caption{(a) A transmission line with non-homogeneous dielectric profile will support forward travelling and backward travelling time-harmonic waves. (b) A transmission line with homogeneous time-varying dielectric profile will support forward travelling and backward travelling space-harmonic waves.}
    \label{fig:1_TL}
\end{figure}
At time $t$ the parameters of the line are switched to $\epsilon_2$ and $\mu_2$. This changes the phase velocity of the wave from $c_1$ to $c_2$, and the characteristic impedance of the line from $Z_1$ to $Z_2$. The continuity of the electric displacement field $\mathbf{D}=\epsilon_i\mathbf{E}_i$ and the magnetic field $\mathbf{B}=\mu_i\mathbf{H}_i$ before and after switching $(i=1,2)$, renders the appearance of a reflected wave, with reflection and transmission coefficients \cite{Shlivinski2018}:
\begin{equation} \label{eq:GammaT}
    \Gamma_s = \frac{1}{2} \Big(\frac{c_2}{c_1}\Big) \Big(\frac{Z_2}{Z_1} + 1 \Big) , \qquad
    T_s      = \frac{1}{2} \Big(\frac{c_2}{c_1}\Big) \Big(\frac{Z_2}{Z_1} - 1 \Big)
\end{equation}
For the reflection and transmission dynamics under soft temporal switching the reader is encouraged to refer to \cite{HadadShlivinsky2020}. In a transmission line with a homogeneous medium that changes with respect to time (see Fig \ref{fig:1_TL}(b)), we can express the voltage at time-point $i$ as: $V_i = V_{0i} \, e^{j(kz - \omega_i t)}$, which is convenient to represent using the voltage phasor: $ V(z,t) = \mbox{Re} \big( V(t) \, e^{jkz} \big) $, where $V(t)$ is a superposition of transmitted and reflected waves: $V(t) = V^+(t) + V^-(t) = V_0^+e^{-j\omega(t) t} + V_0^- e^{j\omega(t) t}$. Thus we define the time-dependant reflection coefficient: $ \Gamma(t) = \frac{V^-(t)}{V^+(t)} $. Upon a small change in the medium parameters, we may apply symmetry considerations and the superposition principle to obtain after switching:
\begin{equation}
    V(t+dt) = \big[ T_s V_0^+ + \Gamma_s V_0^- \big] e^{-j\omega(t) t} + \big[ T_s V_0^- + \Gamma_s V_0^+ \big] e^{j\omega(t) t}
\end{equation}
From this equation we can express the change in $\Gamma$ over a small time interval $dt$:
\begin{equation} \label{eq:gamma_exact}
    \Gamma(t+dt) = \frac{\Gamma_s + \Gamma(t) T_s}{T_s + \Gamma(t) \Gamma_s} e^{2j\omega(t) dt}
\end{equation}
Assuming the change in parameters is small, $T_s \gg \Gamma_s$:
\begin{equation}
    \frac{1}{T_s + \Gamma(t) \Gamma_s} \approx \frac{1}{T_s} \Big( 1 - \Gamma(t) \frac{\Gamma_s}{T_s} \Big)
\end{equation}
And also:
\begin{equation}
    e^{2j\omega dt} \approx 1 + 2j\omega dt
\end{equation}
Substituting in Eq.~(\ref{eq:gamma_exact}) and neglecting second order terms:
\begin{equation}
    \Gamma(t+dt) = \frac{\Gamma_s}{T_s} + \Gamma(t) - \Gamma^2(t) \frac{\Gamma_s}{T_s} + 2j\omega \Gamma(t) dt
\end{equation}
Upon a small change in the impedance of the line we obtain from Eq.~(\ref{eq:GammaT}):
\begin{equation}
    \frac{\Gamma_s}{T_s} = \frac{1}{2} d\ln (Z)
\end{equation}
Finally obtaining:
\begin{equation} \label{eq:riccati_time}
    \frac{d\Gamma}{dt} = 2j\omega \Gamma + \frac{1}{2} (1 - \Gamma^2) \frac{d (\ln Z)}{dt}
\end{equation}
Thus, the behavior of the time-dependent reflection coefficient in a homogeneous transmission line is very similar to the behavior of the space-dependent reflection coefficient in a nonhomogeneous transmission line. In the following, we will use approximations of the governing equations (\ref{eq:riccati_space}) and (\ref{eq:riccati_time}) to reconstruct a dielectric profile (spatial or temporal) from the reflection spectrum and design desired frequency responses.

\subsection{Spatial Profile Reconstruction}

Let's assume a transmission line defined between $z=0$ and $z=L$, connected to a load impedance. The normalized characteristic impedance of the line $Z$ and the propagation constant $\beta$ are a function of space: $ Z = Z(z) \, , \, \beta = \beta(z) $. Riccati's equation (\ref{eq:riccati_space}) that governs the reflection coefficient along the line has no known analytical solution, however, assuming small variations along the line we may assume a small reflection coefficient ($|\Gamma|^2 \ll 1$), obtaining the approximated equation for $\Gamma$:
\begin{equation} \label{eq:riccati_time_approx}
    \frac{d\Gamma}{dz} = 2j\beta \Gamma - \frac{1}{2} \frac{d (\ln Z)}{dz}
\end{equation}
This equation has a closed-form solution. Assuming the line is matched in the end, i.e. $Z(z=L) = Z_L$, the reflection coefficient at the entrance $\Gamma_i = \Gamma(z=0)$ can be formulated according to \cite{Collin}:
\begin{equation}
    \Gamma_i = \frac{1}{2} \int_0^{\theta_L} e^{-j\theta} \frac{d}{d\theta} \ln (Z) \, d\theta
\end{equation}
Where:
\begin{equation}
    \theta = \int_0^z 2\beta(z)dz \qquad , \qquad \theta_L = \int_0^L 2\beta(z)dz
\end{equation}
As observed from the last equation, the normalization of the impedance is arbitrary and does not affect the reflection coefficient. For convenience we choose to normalize by the impedance of the load so: $Z_L = 1$.

Now we assume the material inside the line has permeability $\mu_0$ and an unknown permittivity profile $\epsilon(z)$. In the following we formulate a method to reconstruct the unknown profile $\epsilon(z)$ from the spectrum of the reflection coefficient at the line entrance $\Gamma_i(\omega)$.
Using integration by parts and since $Z \Big|_{\theta = \theta_L} = Z_L = 1$:
\begin{equation} \label{eq:gamma_i}
    \Gamma_i = \frac{1}{2} \Big[ - \ln (Z_0) + j \int_0^{\theta_L} e^{-j\theta} \ln (Z) \, d\theta \Big]
\end{equation}
Where $Z_0$ is the normalized characteristic impedance at $z=0$. We define $\tau = \omega^{-1} \theta$, and through change of variables:
\begin{equation} \label{eq:Gamma_i_omega}
    \Gamma_i(\omega) = \frac{1}{2} \Big[ -\ln (Z_0) + j\omega \int_0^{\tau_L} e^{-j\omega \tau} \ln (Z) \, d\tau \Big]
\end{equation}
Where $\tau_L = 2c_0^{-1} \int_0^L n^{-1}(z') dz'$, $c_0$ is the speed of light in vacuum and $n$ is the refractive index: $n = \sqrt{\epsilon_r}$. We note that the integral on the right side of Eq.~(\ref{eq:Gamma_i_omega}) is the Fourier transform of the natural logarithm of the characteristic impedance (multiplied by a window) with variable $\tau$. Also, the second expression in the right-hand side nullifies when $\omega=0$. Some rearrangement renders:
\begin{equation} \label{eq:GammaSol_w}
    \frac{2[\Gamma_i(\omega) - \Gamma_i(0)]}{j\omega} = \int_0^{\tau_L} e^{-j\omega \tau} \ln (Z) \, d\tau
\end{equation}
Eq.~(\ref{eq:GammaSol_w}) shows that we can obtain the dielectric profile of the line from the spectrum of $\Gamma_i$ by simply using the inverse Fourier transform on its left side. This profile, however, is known in terms of $\tau$, so $\epsilon(z)$ should be estimated from $\epsilon(\tau)$. For that we use:
\begin{equation} \label{dz_dtau}
    dz = (2 c_0 n)^{-1} d\tau
\end{equation}
$z=0$ corresponds to $\tau=0$, then we construct the $z$ axis using this relationship until $Z=L$.

\subsection{Temporal Profile Reconstruction}

Now let us consider a dual problem. An infinite transmission line filled with homogeneous medium with parameters: $\epsilon, \mu$. At time $t=0$, we start changing the parameters of the medium, until at time $T$ the changes stop. We assume the material has only dielectric properties, thus: $\epsilon = \epsilon(t) , 0 < t <T$, $\mu = \mu_0$. As before, for slow changes, we assume $|\Gamma|^2 \ll 1$ and approximate the exact differential equation that governs the reflection coefficient - Eq.~(\ref{eq:riccati_time}) according to:
\begin{equation} \label{eq:riccati_approx}
    \frac{d\Gamma}{dt} = 2j\omega \Gamma + \frac{1}{2} \frac{d (\ln Z)}{dt}
\end{equation}
Here the impedance is normalized by the impedance at time $t=0$. We may follow the same process as before, while defining $\xi= 2c_0 \int_0^t n(t') dt'$ and $\xi_T= 2c_0 \int_0^T n(t') dt'$ and obtain for $\Gamma$ at time $T$:
\begin{equation} \label{eq:GammaSol_k}
    \frac{2[\Gamma_T(k) - \Gamma_T(0)]}{jk} =
    \int_0^{\xi_T} e^{-jk(\xi-\xi_T)} \ln (Z) \, d\xi
\end{equation}
See also \cite{SilbigerHadadMM2022}.
The frequency shift here arises from the initial condition - assuming $\Gamma(t=0) = 0$, meaning there is only a forward propagating wave at $t=0$. Similarly to Eq.~(\ref{eq:GammaSol_w}), that can be used to estimate the spatial profile $\epsilon(z)$ from the spectrum of the reflection coefficient at the entrance $\Gamma_i(\omega)$, Eq.~(\ref{eq:GammaSol_k}) can be used to estimate the temporal profile $\epsilon(t)$ from the k-space measurements of the reflection coefficient at time $T$ - $\Gamma_T(k)$.

\section{Simulation Results}

\begin{figure}[h]
    \includegraphics{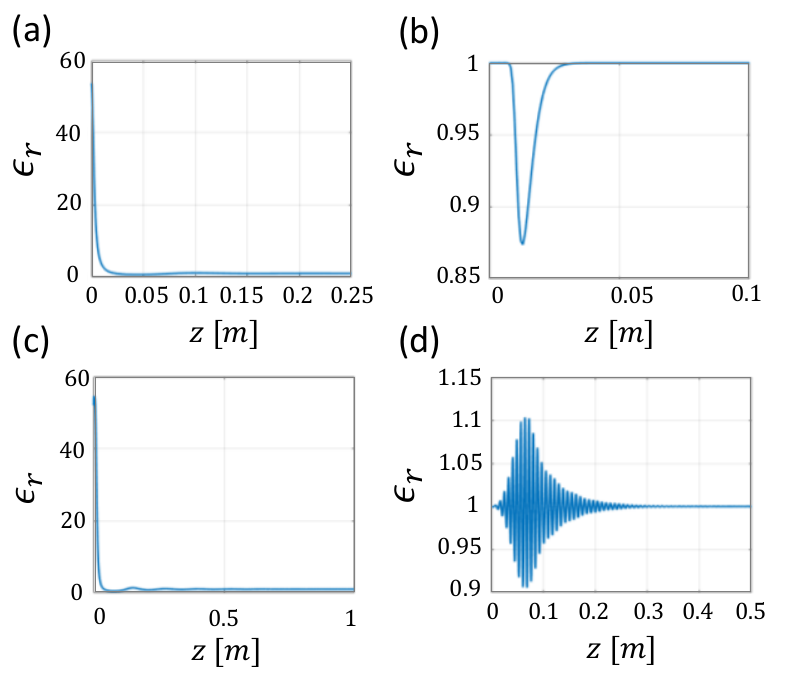}
    \centering
    \caption{Spatial profiles of the dielectric constant $\epsilon_r(z)$ designed to formulate desired frequency responses: (a) Low-pass Chebyshev filter (order 3) (b) temporal differentiator (c) Low-pass Chebyshev filter (order 8) (d) band-pass filter.}
    \label{fig:1z_profiles}
\end{figure}

\begin{figure}
    \includegraphics{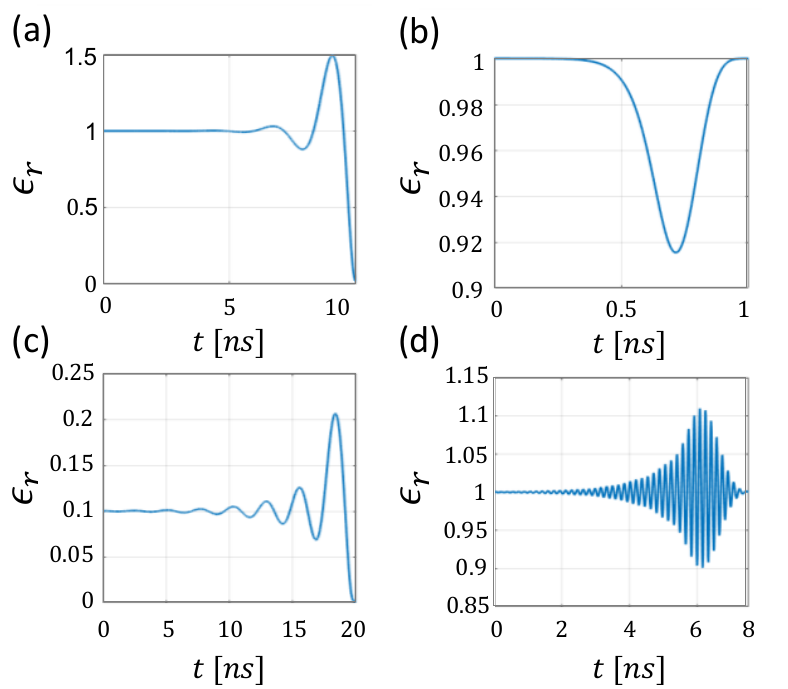}
    \centering
    \caption{temporal profiles of the dielectric constant $\epsilon_r(t)$ designed to formulate desired spatial frequency responses: (a) Low-pass Chebyshev filter (order 3) (b) temporal differentiator (c) Low-pass Chebyshev filter (order 8) (d) band-pass filter.}
    \label{fig:1t_profiles}
\end{figure}

Eqs.~(\ref{eq:GammaSol_w}) and (\ref{eq:GammaSol_k}) allow us to reconstruct the dielectric profile in space (time) from the measurements of the temporal (spatial) reflection spectrum $\Gamma(\omega)$ ($\Gamma(k)$). This is also true if $\Gamma$ is some desired function, and not an actual measurement. Obviously, not every function can be obtained using this scheme, however, this method gives a direct solution that requires no optimization process on the selection of the profile. In the following, we demonstrate how the theory outlined above can be used to construct temporal and spatial analog filters. Here we present four examples - Chebysehv low-pass filter of orders 3 and 8, differentiator and a band-pass filter (see Figs.~\ref{fig:1z_profiles} and \ref{fig:1t_profiles}). For a nonhomogeneous transmission line, the left side of Eq.~(\ref{eq:GammaSol_w}) was calculated from the desired frequency response $\Gamma(\omega)$ to give $\ln(Z(\xi))$. The impedance profile was then inverted, and Eq.~(\ref{dz_dtau}) was used to reconstruct the profile $\epsilon(z)$. It should be noted that Eq.~(\ref{eq:GammaSol_w}) assumes an approximate solution of $\Gamma$, and therefore, the exact reflection coefficient obtained from the reconstructed profile $\epsilon(z)$ will be slightly different from the desired reflection spectrum $\Gamma(\omega)$ that was fed into Eq.~(\ref{eq:GammaSol_w}).

\begin{figure}[h]
    \includegraphics[width=0.5\textwidth]{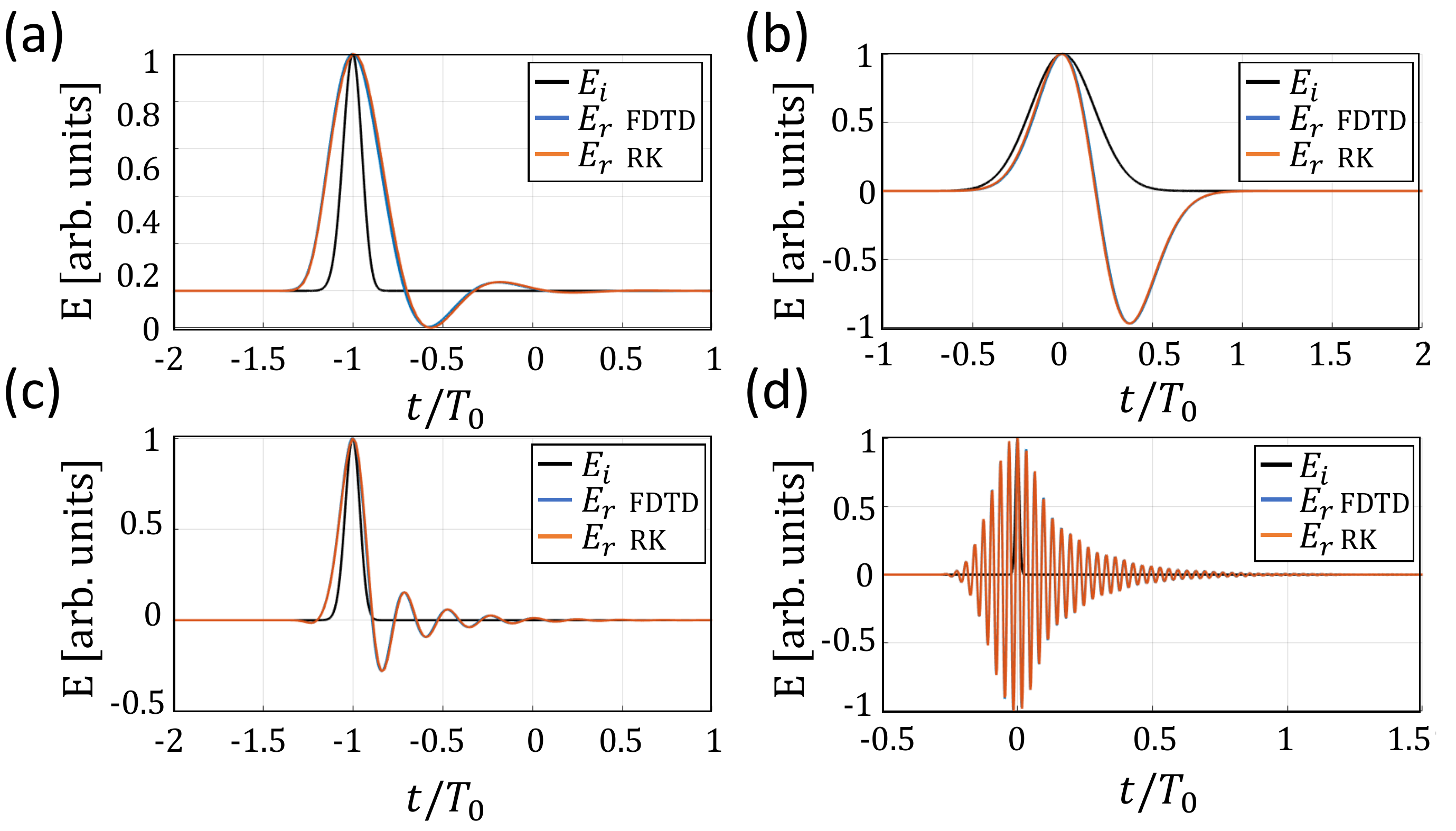}
    \centering
    \caption{Simulation results of a TEM electromagnetic field propagating through a nonhomogeneous medium - here only the electric field is shown. The results in subfigures (a)-(d) correspond to the spatial dielectric profiles in Fig.~\ref{fig:1z_profiles}. (a) Low-pass Chebyshev filter (order 3) (b) temporal differentiator (c) Low-pass Chebyshev filter (order 8) (d) band-pass filter. The time axis is normalized by $T_0=L/c_0$ where $c_0$ is the speed of light in vacuum.}
    \label{fig:2z_fdtd}
\end{figure}

Fig.~\ref{fig:2z_fdtd} shows the reflected fields obtained after a TEM electromagnetic wave propagates through a number of filters, generated by the dielectric spatial profiles in Fig~\ref{fig:1z_profiles}. The incident field is a gaussian pulse (black), and its width is different for each case, to contain the frequencies affected by each filter. The reflected field at the line entrance is calculated in two methods: (1-blue) finite difference time domain (FDTD) simulation of the fields in the line and (2-orange) Runge-Kutta numerical calculation of the reflection coefficient $\Gamma(\omega)$ according to Eq.~(\ref{eq:riccati_time}). The latter method gives the \emph{exact} reflection coefficient, so the predicted spectrum of the reflection coefficient is slightly different from the desired frequency responses that were fed into Eq.~(\ref{eq:GammaSol_w}) to obtain the profiles. The spectrum of the reflected field is then calculated according to: $E^r(\omega) = \Gamma(\omega) E^i(\omega)$, and the temporal profile of the field - $E^r(t)$ is calculated using the inverse Fourier transform. Fig.~\ref{fig:2z_fdtd} shows good agreement of the reflected signal in both methods, suggesting that the frequency response can be tailored with high precision. As the reflections in these examples are small ($|\Gamma| \ll 1$), the amplitudes of the electric fields are normalized.

\begin{figure}[h]
    \includegraphics[width=0.5\textwidth]{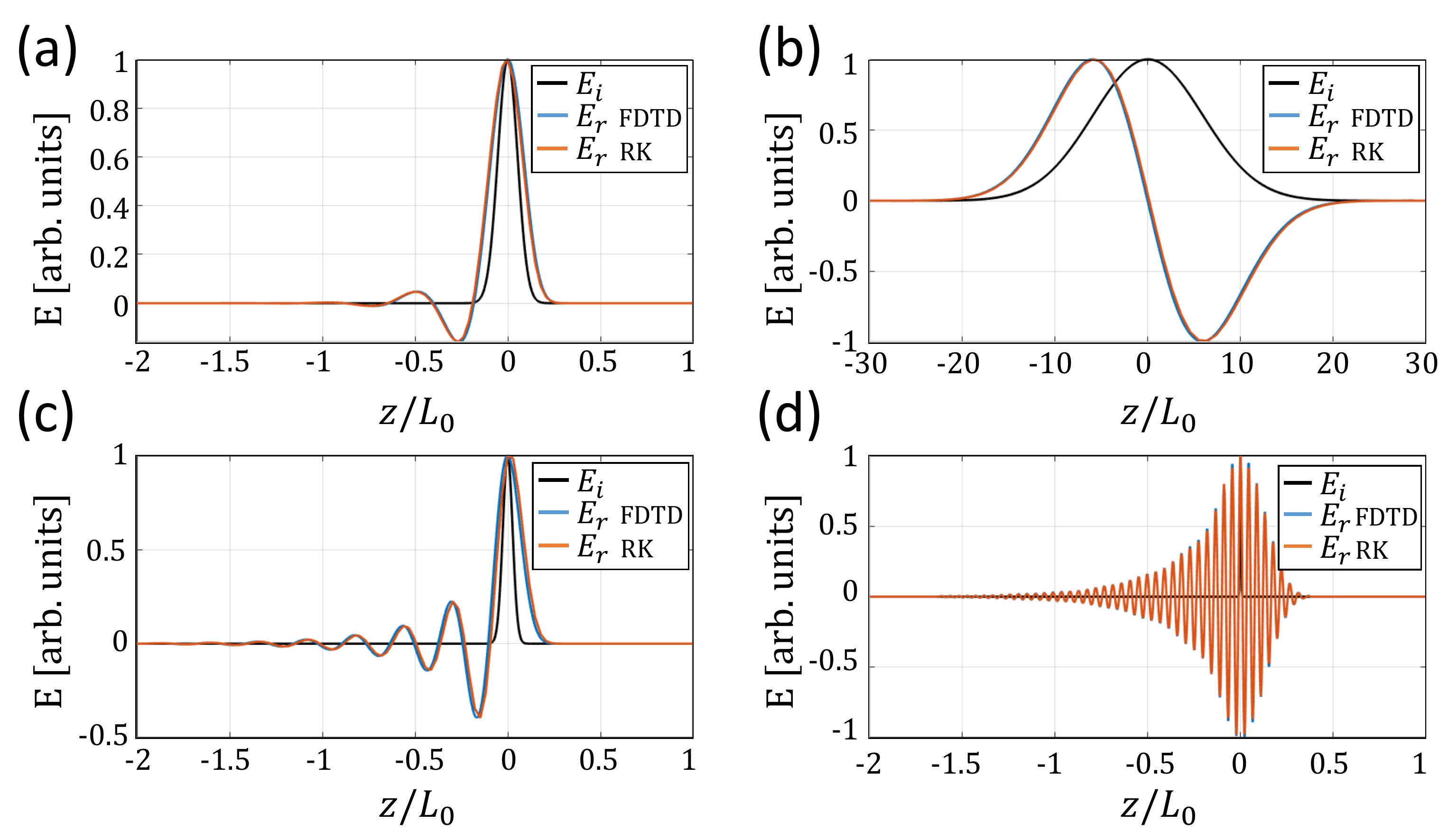}
    \centering
    \caption{Simulation results of a TEM electromagnetic field propagating through a homogeneous medium with temporal switching - here only the electric field is shown. The results in subfigures (a)-(d) correspond to the temporal dielectric profiles in Fig.~\ref{fig:1t_profiles}. (a) Low-pass Chebyshev filter (order 3) (b) temporal differentiator (c) Low-pass Chebyshev filter (order 8) (d) band-pass filter. The $z$ axis is normalized by $L_0=c_0T$ where $c_0$ is the speed of light in vacuum.}
    \label{fig:2t_fdtd}
\end{figure}

To obtain a desired \emph{spatial} reflection spectrum, the process outlined above is repeated using Eq.~(\ref{eq:GammaSol_k}). In the spatially nonhomogeneous medium, at the entrance to the line: $E^t = E^i$, and therefore $\Gamma = E^t/E^i = E^r/E^i$. In this case, however, this relationship is true only for time $t=0$ (and not $t=T$), so we may relate the reflection coefficient only to the transmitted field. However, in a regime of slow and continuous temporal changes, we may assume the reflection is weak compared to the transmission, and with good approximation: $\Gamma(t) \approx E^r(t)/E^i(t)$. Fig.~\ref{fig:2t_fdtd} shows the spatial profile of the reflected electric fields generated by the temporal switching profiles in Fig.~\ref{fig:1t_profiles}. Here we use $E^r(k) = \Gamma(k) E^t(k)$ before using the inverse Fourier transform to obtain $E^r(t)$. In this case as well, the FDTD simulation is in good agreement with the results obtained from the calculation of the reflection spectrum with Eq.~(\ref{eq:riccati_time}).

\section{Conclusions}
In this paper, we have shown the similarity of the scattering problem between a temporally steady-state nonhomogeneous material and a homogeneous, time-dependent material. In both cases, the reflection coefficient is governed by Ricatti's equation, and in both cases it's approximate solution allows a direct solution of the inverse scattering problem for a varying dielectric constant. We have also shown how solving the inverse scattering problem allows the design of desired frequency response, in time (through material non-homogeneities) and space (through temporal switching). This optimization-free technique can be used as a direct method for the design of various analog devices.

After the submission of our manuscript we have encountered \cite{AluHamanyak2022} that was uploaded recently to arXiv and provides another perspective with  analogous results. In our work only the permittivity is time-varying and therefore simultaneously the wave impedance and the wave velocity.

\acknowledgments
This research was supported by the Israel Science
Foundation (grant No. 1353/19).


\end{document}